\documentclass[aps,prl,reprint,showpacs,showkeys,groupedaddress]{revtex4-1}

\usepackage{amsmath, amssymb}
\usepackage{graphicx}
\usepackage{subfigure}
\usepackage{color}
\usepackage{float}
\newcommand*{\rom}[1]{}
\begin{document}


\title{Ratcheting in Nanofluids : Role of confining surface fluctuations}
\author{Aakash}\thanks{aakash.a@students.iiserpune.ac.in}
\author{A. Bhattacharyay}\thanks{a.bhattacharyay@iiserpune.ac.in}
\affiliation{Department of Physics, Indian Institute of Science Education and Research, Pune, Maharashtra 411008, India.}

\date{\today}

\begin{abstract}
We consider the effect of surface undulations of nanochannels on the motion of particles in it. We report the mechanism of surface induced ratcheting transport of particles against fluid flow in nano channels. We show that, the typical number for the velocity of ratcheting particles could be of the order of 50 nm/sec. This is a considerably large effect which could be used for application in efficient filtration of particles smaller than the width of nanopores. 
\end{abstract}

\maketitle


Nanofluidics is the study of  behaviour of fluids confined to nanometer geometry (typically $\sim$ 1 - 100 nm) \cite{kirby2010micro}. Nanofluidics is a very active area of research at present due to the potential of huge applications. Fluids confined to these channels show transport \cite{schoch2008transport} which can be very different from that observed in bulk. Due to small channel sizes which gives control over the flow-through surface fluctuation and modulations, nanofluidics finds its application where fluid samples are handled in very small quantities such as coulter counter \cite{saleh2001quantitative}, chemical and particulate separation and analysis of bio molecules such as DNA, proteins and other ionic species \cite{esmek2019sculpturing,fu2007patterned,zeng2007self,fu2009continuous,napoli2010nanofluidic,wang2008bare,pennathur2005electrokinetic}, cell capture and counting and micropumps to name a few areas. People have applied the principle of nano-transport to perform chromatography with porous medium \cite{saridara2005chromatography,singhal2012separation,asensio2014carbon}. Nanofluidics also finds  application in nano-optics for producing tuneable microlens and for the detection of viruses and nanoparticles \cite{mitra2010nano}. Nanofluidics is also significant in the area of biotechnology and molecular diagnosis. \cite{jayamohan2013applications}. Some carbon nanotubes (CNT) can give rise to ultra-efficient transport mechanism for water and gas molecules \cite{noy2007nanofluidics,tao2018confinement,sparreboom2010transport}. Water transport rate in these nanotubes can be far exceeding that predicted by theory. CNT (Carbon nanotubes) finds their application in desalination and purification of water. Some carbon nanotubes (CNT) exhibit remarkable electrical, thermal and mechanical properties like tensile strength etc. \cite{tans1997individual,berber2000unusually,kim2001thermal}. Apart from these there is a natural analogy between fluid flow in nanochannels and electron and holes motion in semiconductors electronic devices. And this analogy can be exploited in electronic functioning such as rectification of current and bipolar and field effect transistor actions \cite{perdigones2014correspondence}. 
\par
In the present paper we are interested in exploiting the possibility of surface driven ratcheting \cite{ait2003brownian,ethier2019tilted} in a nano-channel. Ratcheting involves transport of particles against a barrier potential being driven by a directionless force in the presence of thermal noise. The process generally involves a non-equilibrium transport of particles against a force. An essential ingredient for any transport process is inversion symmetry breaking. Normally, in the paradigmatic models of ratcheting of a particle one considers this inversion symmetry being broken at small length scales. This broken inversion symmetry in the presence of thermal noise and a non directional drive to intermittently force the system out of equilibrium essentially results in directed motion of particle.
\par
A recent paper by Marbach et al. \cite{marbach2018transport} reports an important general method of analyzing particle transport through nano-channels in the presence of fluctuations of the channel wall. This work develops a very essential tool of spectral analysis of motion of test particles in a nano-channel. Under lubrication approximation, this work captures diffusivity renormalisation of a test particle in a nano-channel under various surface fluctuations of the system. In the present paper, in what follows, we will be directly using the relation derived by Marbach et al. \cite{marbach2018transport}, to show that, the overdamped velocity dynamics of the test particle under surface fluctuations indicates that there exists possibility of a surface fluctuation driven ratcheting of particles in such systems.
\par
Surface driven ratcheting of particles in nanopores can result in efficient filtration because one can make the particles move directionally against the flow of the fluid. This would be equivalent to the ratcheting of a particle up against a potential barrier. And, this is exactly where lies the importance of this phenomenon in the realm of nanofluidics by employing which one can think of general mechanism of controlled filtration. Moreover, the existence of such a general mechanism in the realms of nanofluidics also indicates that the mechanism may be already in use in biological systems where nanofluidcs is a norm than an exception.
\par
The way a paradigmatic ratcheting particle moves against a potential, the same mechanism can drive a diffusing particle in a nanofluid to move against the flow of the fluid \cite{astumian1994fluctuation,astumian1997thermodynamics,bhattacharyay2012directed,tarlie1998optimal,cisne2011particle,loutherback2009deterministic,celestino2011generic}. The very essential inversion symmetry breaking at mesoscopic scales can possibly be realized in nanofluidics by patterning the elasticity of the surface of the confinement of the fluid. By taking a general approach of using a sawtooth surface undulations in the presence of temporal oscillations we show in this paper that surface driven ratcheting can happen and a diffusing particles being subject to such surface fluctuations can be driven against the flow of the fluid in a nano-channel without applying any directional or convective force. 
\par
We organize the paper in the following way. For completeness we first present the salient steps of the calculations by Marbach et al. \cite{marbach2018transport} deriving the general velocity dynamics of the test particle in a nano-channel under surface fluctuations in the lubrication approximations. Then we show how this overdamped dynamics results in a ratcheting in the presence of an inversion symmetry broken patterning of the surface. We show numerical results of this ratcheting process. Following which we conclude the paper with a detailed discussion of the model.  
\section{Marginal Advection - Diffusion equation}
\par
In the following we present the basic analysis as done by Marbach et. al. in ref. \cite{marbach2018transport} to setup the stage for the analysis of ratcheting. The dynamics of probability density $p(\bf{r},\emph{t})$ of a tracer particle in the presence of velocity field $\bf{u}$($\bf{r}$,\emph{t}) is governed by Fokker-Planck equation. 

\begin{equation}
\frac{\partial p({\bf r},t)}{\partial t} = D_{0} \nabla^{2}p({\bf r},t) - \nabla\cdot[{\bf u}({\bf r},t) p({\bf r},t)] 
\end{equation}
\par
Let $h_{u}$({\bf x},\emph{t}) be the upper surface profile, where {\bf x} = ($x$, $y$) is the coordinate of the 2D plane at a fixed height $z$ and $h_{u}(\bf{x},\emph{t})$ = $H + h({\bf x},\emph{t})$. Where $H$ is the mean height of an upper fluctuating surface and $h({\bf x},t)$ is the height variation over the mean height. In this calculation \cite{marbach2018transport}, the lower surface confining the fluid layer is considered to be fixed without any loss of generality. Therefore $z$ lies between 0 and $h_{u}$({\bf x},\emph{t}). The marginal probability distribution function $p^{*} ({\bf x},t)$ is then defined as 
\begin{equation}
p^{*} ({\bf x},t) = \int_{0}^{h_{u}({\bf x},t)}  dz \: p({\bf r},t)
\end{equation}
\par
Now integrating both sides of Fokker-Planck eqn. (1) from 0 to $h_{u}$({\bf x},\emph{t}) while using the Leibnitz rule and then simplifying, one obtains an equation for marginal probability distribution
\begin{equation}
    \frac{\partial p^{*}({\bf x},t)}{\partial t} = D_{0} \nabla^{2}_{||}p^{*}({\bf x},t) - \nabla_{||}\cdot[{\bf v}({\bf x},t)p^{*}({\bf x},t)]
\end{equation}
where 
\begin{equation}
    \nabla_{||} = \hat{i}\frac{\partial}{\partial x}+\hat{j} \frac{\partial}{\partial y}
\end{equation}
and
\begin{equation}
\nabla_{||}^{2} = \frac{\partial^{2}}{\partial x^{2}} + \frac{\partial^{2}}{\partial y^{2}}.
\end{equation}
\begin{multline}
    {\bf v}({\bf x},t) = D_{0} \frac{\nabla_{||}h({\bf x},t)}{h_{u}({\bf x},t)} + \frac{1}{h_{u}({\bf x},t)} \int_{0}^{h_{u}({\bf x},t)} dz \: {\bf u_{||}} ({\bf r},t) 
\end{multline}    
is the effective velocity field that a tracer particle is getting to see. The equation (6) for effective velocity will be important for later discussions. In deriving the equation (6) one makes use of the following boundary conditions \cite{marbach2018transport}.
\par
\begin{enumerate}
\item Kinematic Boundary condition : It implies that the velocity of the fluid at the boundary is zero with respect to the boundary.
\begin{multline}
    u_{z}({\bf x},h_{u}({\bf x},t),t) - \frac{\partial}{\partial t} h_{u}({\bf x},t) - {\bf u}_{||}({\bf x},h_{u}({\bf x},t),t).\\  \nabla_{||}h_{u}({\bf x},t) = 0
\end{multline}
Where ($\nabla_{||}, \frac{\partial}{\partial z}$) and (${\bf u}_{||}, u_{z}$) denote in plane and out of plane coordinates respectively.

\item Conservation of probability at z = 0 :
\begin{equation}
    \frac{\partial}{\partial z} p({\bf x},z=0) = 0
\end{equation}

\item  Conservation of probability at $z = h_{u}({\bf x},t)$ :
\begin{equation}
    \nabla p({\bf x},z)|_{z = h_{u}({\bf x},z)}\cdot{\bf n} = 0
\end{equation}
\end{enumerate}
\par
Using equation (3) and above three boundary conditions one obtains the following equation for marginal probability density $p^{*}({\bf x},t)$
\begin{multline}
    \frac{\partial}{\partial t}p^{*}({\bf x},t) = D_{0}\nabla^{2}_{||}p^{*}({\bf x},t) - D_{0}\nabla_{||}\cdot[p({\bf x},h_{u}({\bf x},t),t)\\\nabla_{||} h_{u}({\bf x},t)] - \nabla_{||}\cdot\int_{0}^{h_{u}({\bf x},t)} dz \: {\bf u_{||}} \: ({\bf r},t) \: p({\bf r},t)
\end{multline}
\par
But this equation is not in terms of marginal probability density $p^{*}({\bf x},t)$ only. It contains terms involving $p({\bf r},t)$ also. However if one makes use of the Lubrication approximation as is shown by Marbach et al., then one can assume the equilibrium in $z$ direction and factorize the $p({\bf x},z,t)$ into $f({\bf x},t)$ and some function of $z$. Then using the normalisation one can write $p({\bf x},z,t) \approx f({\bf x},t)$. This equation greatly simplifies the above equation and one gets 
\begin{multline}
    p^{*}({\bf x},t) \approx \int_{0}^{h_{u}({\bf x},t)} dz \: f({\bf x},t) = h_{u}({\bf x},t) f({\bf x},t)
\end{multline}
\par
Therefore we have the following equation for marginal probability density $p^{*}({\bf x},t)$
\begin{multline}
    \frac{\partial}{\partial t}p^{*}({\bf x},t) = D_{0}\nabla^{2}_{||}p^{*}({\bf x},t) - D_{0} \nabla_{||}\cdot[p({\bf x},h_{u}({\bf x},t),t)\\ \nabla_{||}h_{u}({\bf x},t)] - \nabla_{||}\cdot[p^{*}({\bf x},t)\overline{{\bf u_{||}}({\bf x},t)}]
\end{multline}
Where $\overline{{\bf u_{||}}({\bf x},t)}$ is average velocity defined by :
\begin{equation}
    \overline{{\bf u_{||}}({\bf x},t)} = \frac{1}{h_{u}({\bf x},t)} \int_{0}^{h_{u}({\bf x},t)} dz \: {\bf u_{||}}({\bf x},z,t)
\end{equation}
\par
Finally the following equation is obtained for marginal probability density 
\begin{equation}
    \frac{\partial p^{*}({\bf x},t)}{\partial t} = D_{0} \nabla^{2}_{||}p^{*}({\bf x},t) - \nabla_{||}.[{\bf v}({\bf x},t)p^{*}({\bf x},t)]
\end{equation}
where 
\begin{equation}
    {\bf v}({\bf x},t) = D_{0} \frac{\nabla_{||}h({\bf x},t)}{h_{u}({\bf x},t)} + \frac{1}{h_{u}({\bf x},t)} \int_{0}^{h_{u}({\bf x},t)} dz \: {\bf u_{||}} ({\bf r},t) 
\end{equation}    
is the effective velocity field which the tracer particle is moving with.

\section{Surface induced ratcheting}
Let us have a look at how the surface fluctuations induced ratcheting can happen in nano-fluid systems. Under the lubrication approximation, the velocity of the diffusing particle as arrived at by Marbach et al., is 

\begin{equation}
    \frac{dx}{dt} = D_{0} \frac{\frac{\partial}{\partial x}h(x,t)}{H + h(x,t)} + \frac{1}{H + h(x,t)} \int_{0}^{H+h(x,t)}dz \: u(x,z,t)
\end{equation}
where we have considered only one of the planar coordinates ($x$,$y$) for the sake of simplicity.
In the above equation, there are two terms on the RHS of which the first term gives the average velocity induced by diffusion and surface gradients and the second term stands for the advection of the particle by the fluid of velocity $u(x,z,t)$. If the first term can produce a velocity which is  opposite in direction to the second term and can overcome the advection then the particle can actually move in opposite direction to the fluid flow resulting in enhanced filtration. As it is obvious from the presence of gradient term $\frac{\partial}{\partial x}h(x,t)$, the sign of the velocity due to this first term can be reversed and we can always get this part of velocity to be opposite to that coming from the second term. However, this will require a breaking of inversion symmetry at smaller scales.
\par
To look at the possibility of ratcheting in a simpler way, let us consider the velocity of the particle coming from the diffusive coupling alone in the absence of any noise term which would be shown in the later part of the paper to play a very useful role. This gives the velocity of the particle to  be 
\begin{equation}
    \frac{dx}{dt} = \frac{D_{0}}{H} \frac{\partial}{\partial x}h(x,t)
\end{equation}
Where we have gone by the assumption $H \gg h(x,t)$ which is the lubrication approximation. This equation is the overdamped dynamics of a particle in the potential $h(x,t)$. To have a non-trivial result let us consider that $h(x,t)$ = $\chi(x) T(t)$ such that there exists no traveling modes in the surface fluctuations. The nontrivial result of ratcheting should appear in the absence of any traveling mode in the fluctuations of the surface height which plays the role of the effective potential here. Consider the profile of $\chi(x)$ and $T(t)$ as shown in fig.1 and 2.
\begin{figure}[ht]
  \includegraphics[width=0.5\textwidth]{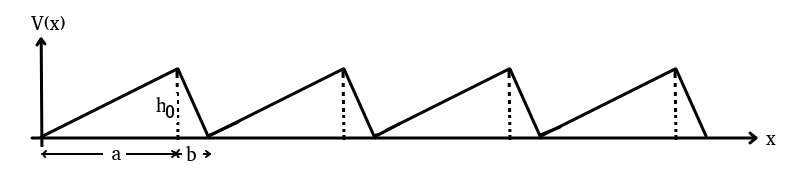}
\caption{Spatial part of potential, sawtooth  potential with period $a+b$ (showing the broken symmetry)}
\label{fig:1}       
\end{figure}
\begin{figure}[ht]
  \includegraphics[width=0.5\textwidth]{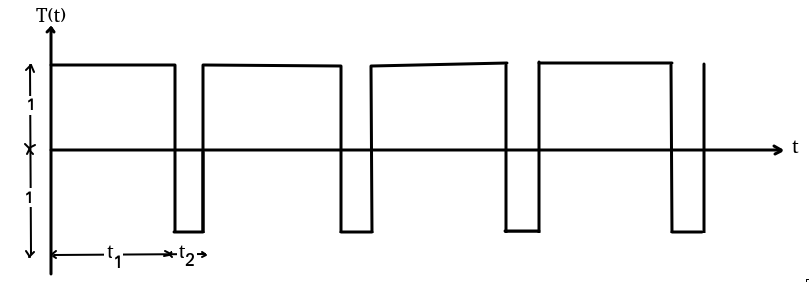}
\caption{temporal part of potential with period $t_{1}+t_{2}$ (with positive and negative pulse)}
\label{fig:2}       
\end{figure}
\begin{figure}[ht]
  \includegraphics[width=0.5\textwidth]{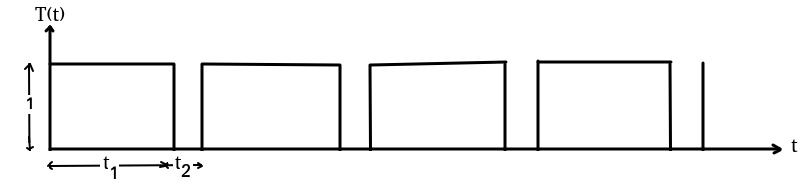}
\caption{temporal part of potential with period $t_{1}+t_{2}$ (with positive and zero pulse)}
\label{fig:3}       
\end{figure}
\par
We are taking $\chi(x)$ to have a sawtooth profile to have the essential ingredient of ratcheting process which is an inversion symmetric breaking of space at small scales. This can be achieved by patterning the surface of the nano-channel in many ways keeping in mind that what one needs is different lengths of differing elasticity on the surface in a periodic manner to have such inversion symmetry breaking under the temporal non-traveling fluctuations on the surface. 
\par
The sawtooth potential in this particular way of modelling plays an essential role which is worth mentioning. It is a very standard practice in the existing literature to consider the inversion symmetry breaking potential of sawtooth form. It is required keeping in mind that we are actually modelling the system by an overdamped dynamics in the absence of a noise term in the beginning. The average effect of the noise on transport is already taken into account in the diffusivity $D_{0}$. Thus, if $\chi(x)$ is a sawtooth function of space, in this way of modelling, the particle will not take infinite time to reach the maximum of $\chi (x)$ in the absence of the fluctuations. Having a discontinuity at the maximum and minimum of the function $\chi(x)$ rids one of this stickiness of the equilibrium points which, otherwise, will have a slowdown effect as the particle reaches the maximum of a smooth $\chi(x)$ function. In a real system, this problem of sticking, even in the case of a smooth $h(x)$ profile, will not be there in the presence of thermal fluctuations.
\par
Having taken into consideration the broken inversion symmetry by $\chi(x)$ in this way which can in principle be realized by surface patterning, let us pay attention to the other most important ingredient for ratcheting and that is intermittently disturbing the system to drive it out of equilibrium which is achieved by $T(t)$. The periodic forcing $T(t)$ can have time period $t_{1}+t_{2}$ where over the time $t_{2}$ this forcing induces a change in sign of the $\chi(x)$ profile and the investigation of the ratcheting using the present model crucially depends on the choice of this intermittent forcing from which the system (the particle) gains energy to overcome the barrier.
\par
Let us qualitatively understand the part of the ratcheting phenomenon in the absence of noise in this model explicitly. As is shown in fig 4.
\begin{figure}[ht]
  \includegraphics[width=0.5\textwidth]{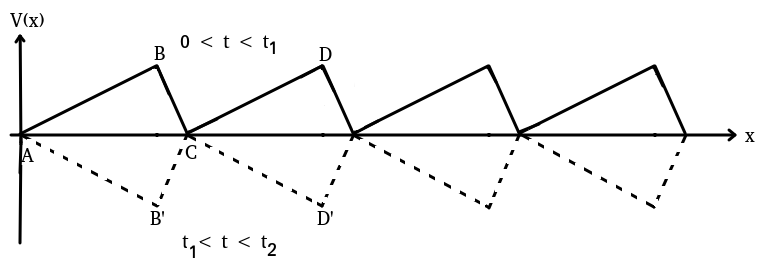}
\caption{}
\label{fig:4}       
\end{figure}
Consider that the particle was initially sitting at the position A when $T(t)$ is positive. Under the action of the force generated by $\chi(x)$ the particle will move to the point B and will stay in the vicinity of point B for all the time if $T(t)$ does not change its sign. Now consider a flip in sign of the $T(t)$ making an inversion of $\chi(x)$ which brings the particle at B$^{\prime}$. Here the particle again becomes unstable and will start moving either towards C or A. Now, if the interval $t_{2}$  is chosen such that the particle can have about just enough time to reach C and not A then when $T(t)$ becomes positive again, the particle will either be at C or somewhere between the points A and B. If the particle is somewhere between A and B then it will come back to B. However, if the particle is at C then it can again move both ways and can either end up at B or at D. The choice of time interval $t_2$ will play a role here in which way the particle takes i.e., towards B or D in the absence of fluctuations where fluctuations can help it overcome this stricter dependence of transport on $t_2$. However, there definitely exist a finite probability of  the particle to go to D and thus, there can happen ratcheting. This ratcheting is just a consequence of the theory developed by Marbach et al., where directed transport can happen without any directional driving. The consideration of noise on top of this will make the process easier by removing the stricter control over the time scale $t_2$.
\par
In any flashing ratcheting model the turning of this on-off time of the intermittent force crucially determines the efficiency of the ratcheting and the present model is no exception. In numerically evolving the system, we have kept in mind the crucial choice of $t_{2}$ time scale and show our numerical results for the deterministic as well as more realistic stochastic situation. In the stochastic case we add a stochastic term to the model which is a Gaussian white noise of strength $\sqrt{2D_{0}}$ which modifies the governing equation as :
\begin{equation}
    \frac{dx}{dt} = \frac{D_{0}}{H} \frac{\partial}{\partial x}h(x,t) + \sqrt{2 D_{0}} \xi (t)
\end{equation}
Where $h(x,t)$ = $\chi(x)T(t)$ and  $\xi(t)$ is Gaussian white noise with zero mean and unit standard deviation.

\section{Numerical Results}
\par
In the Fig.5 we show the directed transport results from the direct simulation of eqn.(17) where the noise is not taken into account. The choice of the parameters are $a$ = 4 nm, $b$ = 1 nm, and $\: h_{0}$ = 1 nm. The time step of simulation is $\Delta t = 0.01 $ and we have employed fourth order Runge-Kutta method. 
\par
The diffusivity $D_0 = 10^4$ $\text{nm}^2$/sec and the radius of the channel $H = 100$ nm are set. The velocity scale that $D_0/H$ sets is 100 nm/sec. The typical diffusivity of a protein monomer at room temperature in water is about $10^{8}$ $\text {nm}^2$/sec. The size of the molecule that we are considering here is roughly larger than $10^3$ times in diameter than that of a protein monomer. Our simulation shows that molecules of this size or larger would actually show ratcheting. For smaller molecules, diffusion will dominate.
\par
Figure 5 shows that, there is a dependence of the velocity (here the slope of the trajectory) of ratcheting on $t_2$ as was expected. Actually, we got it to be maximum, for the given set of parameters, at $t_2=0.115$ sec. For this range of other parameters, the ratcheting actually happens over a very small range of $t_2$ around 0.1 sec. Beyond the above mentioned range of $t_2$ the particle practically gets stalled. The choice of $t_1$ is quite flexible so long it is larger than $t_2$ and can be kept even very large which would then reduce the velocity of ratcheting.

\begin{figure}[ht]
  \includegraphics[width=9.0cm,height=6cm]{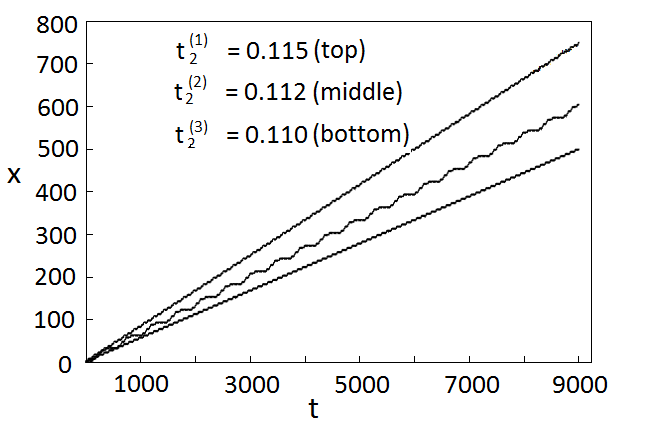}
\caption{Evolution for deterministic case($x$ is in the unit of nm and t is in the unit of second)}
\label{fig:5}       
\end{figure}

\par
It is seen that addition of noise makes the model quite robust and increases the velocity of ratcheting by a couple of orders of magnitude. The ratcheting happens not just in the narrow band of $t_2$, as it was for the case without noise. Ratcheting happens for practically all value of $t_{2}$ parameter when we simulate the stochastic eqn.(18). We show these results in the fig.6. As shown in fig.6, although ratcheting happens over a wide range $t_2$ the velocity of ratcheting is highly sensitive to the change in $t_{2}$. 
\begin{figure}[H]

\includegraphics[width=9.5cm,height=6.0cm]{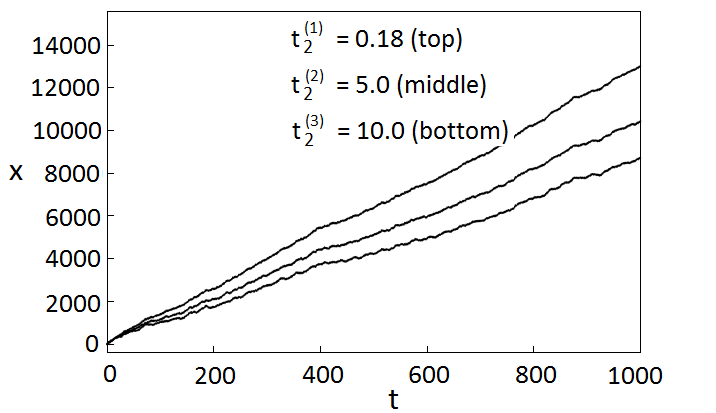}
\caption{Evolution for stochastic case ($x$ is in the unit of nm and $t$ is in the unit of second)}
\label{fig:6}       
\end{figure}
\par
 In the same way we also considered and obtain the results for the case when time pulse oscillates between positive value and zero (fig.3). We have obtained the same qualitative and almost the same quantitative behaviour for $x-t$ curves.
 
\par
 To check the dependence of the velocity of ratcheting on the other relevant parameter $h_0$, we set $t_2$ a constant ($t_2=1.0$) and vary $h_0$. We plot velocity of ratcheting against $h_0$ in fig.7 to have an idea of how the amplitude of surface oscillations can influence ratcheting. In this plot, parameters $a$ and $b$ are kept at the same value as in other ones. The velocity does rise appreciably as $h_0$ varies from 1 to 8 and then falls. We cannot increase $h_0$ roughly beyond 10 nm given the fact that $H = 100$ nm and we are under lubrication approximation. So, neither very small nor very large amplitude of surface oscillations will help increase ratcheting.
 
\begin{figure}[ht]
\includegraphics[width=9.0cm,height=7.0cm]{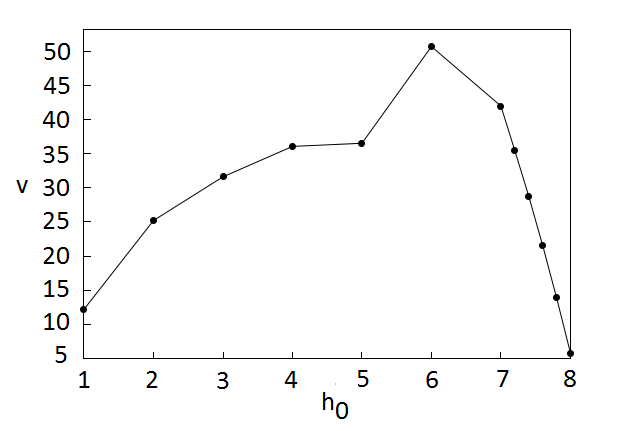}
\caption{Effect of surface undulation amplitude $h_{0}$ on ratcheting velocity $v$ ($v$ is in nm/sec and $h_{0}$ is in nm.) }
\label{fig:7}       
\end{figure}

\section{discussion}
We have shown in this paper that, based on a broader framework as established by Marbach et al., \cite{marbach2018transport} there can happen surface fluctuations driven ratcheting of Brownian particles in a nanofluidic channel. This ratcheting of particles can result in velocity of small particles against the motion of a fluid. This in principle can result in effective filtration of particles which are smaller than the pore size. 
\par 
The typical velocity of the ratcheting that we get is of the order of tens of nanometer per second. This is a considerable velocity that the boundary induced ratcheting can result in where the tube width is about 100 nm. This velocity will increase with the decrease of the channel width, however, that possibly is a regime falling outside lubrication approximation. This means that the velocity of the fluid against which particles will move can safely be about 10 nm/sec. 
\par
This mechanism of ratcheting which is possible in nano-fluids under surface fluctuations is a generic one. Most probably such a mechanism is in use in some biological systems where particle transport happens without any directional drive. Effects of other finer details on such transport can be addressed once the existence of the basic mechanism is established. We hope that, possibility of the application of this mechanism in designing efficient filters is quite possible given the ability of nano-scale engineering of such channels these days.   
\par
The present analysis of the generic mechanism critically depends on the general revelation by Marbach et al., \cite{marbach2018transport} of the coupling of diffusivity of particles to the surface fluctuations. In the present paper we have just utilized this coupling which can practically happen in the domain of nanofluidics. In nano-scale flow the typical velocity of fluid through these channels are not very high and the advection term can be comparable to the ratcheting velocity. Moreover, the fluctuation scales in such systems would also be of the order of the particle size within these channels. Matching of all these scales is the key for the emergence of ratcheting induced filtration in nano-channels.

\bibliography{references.bib}
\end{document}